# Constant TMR magnetic field sensor detectivity with bias voltage

E. Monteblanco[1], A. Solignac[1], C. Chopin[1], J. Moulin[1], P. Belliot[2], N. Belin[2], P. Campiglio[2], C. Fermon[1], M. Pannetier-Lecoeur[1]

[1] SPEC, CEA, CNRS, Université Paris-Saclay, CEA Saclay 91191 Gif-sur-Yvette Cedex, France

[2] CrivaSense Technologies SAS, 91190 Saint Aubin, France

In this letter, we present a study of optimized TMR magnetic field sensors as a function of voltage bias. The 1/f low-frequency noise is quantified by the Hooge-like parameter $\alpha$ which allows to compare the low-frequency behavior of various TMR sensors. The sensitivity as well as the detectivity of the sensor are characterized in the parallel state and at 0 mT. We observe that the sensitivity shows a strong voltage dependence and the noise presents an unexpected decrease, not anticipated by the Hooge's law. Moreover, surprisingly, an almost stable detectivity (140-200 nT/√Hz at 10 Hz and 15-20 nT/√Hz at 1 kHz) as a function of the bias voltage is observed, tending to highlight that the variation of sensitivity and noise are correlated. Even if the I-V curves are strongly non-linear and reflect the different symmetries of the conduction bands channels, the variations in sensitivity and noise seems to depend mainly on the distortion of the MgO barrier due to bias voltage. With a simple model where the normal noise and sensitivity of the TMR sensors are modified by an element having no noise and a parabolic conductance with voltage, we describe the behavior of noise and sensitivity from mV to V.

## Introduction

Technological devices based on magnetic tunnel junctions (MTJs) such as non-volatile memories (MRAM)[1], spintronic nanoscillators[2] (STO), neuromorphic computing[3], and magnetic sensing for industrial applications (automotive or consumer[4]) are in continuous evolution. For magnetic sensing, the interest of devices based on MTJs is their very high tunnel magnetoresistance (TMR) ratio which induces a high voltage output proportional to the bias voltage. In order to increase the performances in terms of signal to noise ratio, large arrays of TMR pillars in series and parallel or hybrid devices are used[5-7]. This allows to maintain a high TMR and to reduce the low frequency noise. To apprehend the physics and functioning of the possible varieties of devices, it is necessary to understand and model the response of the individual elements composing the device, regardless of the number integrated afterwards. In addition, the current trend is to reduce the active area of TMR sensors, typically by reducing the number of individual elements in series while maintaining a high output voltage. Understanding and modelling the behaviour of individual TMR pillar as a function of the applied bias voltage is therefore also necessary in this case.

The sensor properties versus bias voltage are then of high importance especially because of the non-linear behaviour of the I-V sensor response. Several studies have been performed before on the MTJ conductance which allow to access to the complex conduction band properties in MgO tunnel barrier and interfaces [8-11]. Nevertheless for applications, the sensors performances in terms of sensitivity, noise and detectivity need to be studied as a function of the bias voltage and in a low-frequency range.

In this paper we present a complete study of the 1/f low-frequency noise, which is the main source of noise in magnetic field sensors based on MTJ. Especially, we observe an unexpected constant detectivity versus the bias voltage. A simple model is proposed to explain the noise, sensitivity and detectivity behaviour as a function of the bias voltage. The model is validated over several TMR sensor structures and over 2 orders of magnitude in bias voltage.

## Sample studies

MTJ systems are basically composed by two ferromagnetic layers separated by an insulating layer. The reference layer magnetization is blocked into one direction (sensitivity axis), while a free layer magnetization rotates following the direction of the magnetic field to be detected, inducing a variation of the structure resistance due to TMR spintronic effect[10]. The multilayer stack 1 of the TMR sensor studied in this manuscript has the following structure: $SiO_2$ / Ta(5) / $SyF_1$ / MgO / $SyF_2$ / Ta(10) (thicknesses in nanometers), where the synthetic ferrimagnet $SyF_1$ is composed by a multilayer PtMn(25) / CoFe(2.3) / Ru(0.83) / CoFe(1) / Ta(0.1) / CoFeB(1.5) and corresponds to the reference layer thanks to the blocking due to the strong RKKY coupling and exchange bias[12]. The free $SyF_2$ layer is composed by CoFeB(1.5) / Ta(0.1) / CoFe(1) / Ru(2.6) / CoFe(2) / PtMn(16) and corresponds to a pinned free layer[13].

The samples were first post annealed at 300 °C during 30 min with a magnetic field of 1 T to pin the reference $SyF_1$ magnetization into the sensitivity axis direction (0°) and to obtain a good crystallization of the CoFeB- barrier interfaces. A second annealing was done at 300 °C during 30 min with a magnetic field of 80 mT to fix the $SyF_2$ layer magnetization at 90 ° from the reference sensitivity axis. Samples were processed after the annealing using standard optical UV lithography and ion milling. Several types of arrays are studied with different number of pillars N and the diameter D (µm) but with a resistance area (RA) product by pillar at 0 mT around 15 kΩ.µm² (stack 1). Samples were wire bonded to a chip for the transport and noise measurements at room temperature. The TMR for all devices is between 150-200 % at room temperature for a voltage bias of 100 mV. This highlights the good quality of the MgO barrier, a good crystallographic matching and an effective electronic filtering[14]. The orthogonal magnetization positions of both SyF layers allows to linearize the magnetic response of the TMR sensor when the magnetic field is applied along the sensitivity axis direction[15], as can be seen with the black MR curve on Figure 1(a). Parallel (P) and antiparallel (AP) magnetic configuration[8, 11] is translated as lower and higher resistance in this curve and the $SyF_1$ spin-flop regime appears overcoming + 50 mT. In all cases geometrical and structural parameters have been optimized to reduce the hysteresis of the curve, remaining linear around zero field, as it is expected by sensors applications[16, 17].

In order to remain below the breakdown voltage or to avoid a possible degradation of the MgO barrier, the maximum $V_{pillar}$ applied per TMR pillar is around 1 V. In order to characterize correctly the 1/f noise from the TMR sensor, the setup is installed in a mu-metal magnetically shielded room. A battery is used to provide a stable bias voltage to the devices through a balanced ¼ Wheatstone bridge (WB). The output signal of the bridge is amplified using an INA103 low-pass amplifier (gain 489) and then amplified again (gain 10) and bandpass filtered (0.1-3 kHz). In this range, the 1/f noise is dominating. Temporal signals are acquired and a fast Fourier transform (FFT) is used to measure the noise spectral density (average of 30 times). For measuring the sensitivity of our devices, an alternating external calibrated magnetic field ($H^{rms}_{8.5\,\mu T}$) of 8.5 µT$^{rms}$ at 30 Hz was applied along the sensitivity axis, so using the output signal at 30 Hz (peak, $V^{rms}_{30Hz}$), we can calibrate the sensor.

### Definition of noise, sensitivity and detectivity

From the noise amplitude spectral density $\sqrt{S_V}$ and the output signal at 30 Hz ($V^{rms}_{30Hz}$) measured from a balanced ¼ Wheatstone bridge, we obtain the Hooge-like parameter $\alpha$ to quantify the 1/f low-frequency noise, the sensitivity and the detectivity of the TMR sensor from the following equations:

$$\alpha = \frac{S_V \cdot N \cdot A \cdot f}{V^2_{TMR}} \quad (1)$$

$$Sensitivity\ \left[\frac{\%}{T}\right] = 100 \cdot \frac{V^{rms}_{30\,Hz}}{H^{rms}_{8.5\,\mu T} V_{TMR}} \quad (2)$$

$$Detectivity\ \left[\frac{T}{\sqrt{Hz}}\right] = \frac{\sqrt{S_V}}{Sensitivity \cdot V_{TMR}} = \sqrt{\frac{\alpha}{NAf}} \frac{H^{rms}_{8.5\,\mu T} V_{TMR}}{V^{rms}_{30\,Hz}} \quad (3)$$

where $A$ is the transversal area of a MTJ pillar, N is the number of pillars in series per TMR, $f$ is the frequency, $V_{TMR}$ is the bias voltage of a TMR sensor which is $N$ times the voltage of each pillar $V_{TMR} = N \cdot V_{pillar}$ composing the TMR. The detectivity is the field for which the signal to noise ratio is 1. It corresponds to the smaller value of magnetic field that the sensor can detect. All these performances were corrected using an electronic WB circuit correction factor[16].

### Experimental results

The noise amplitude spectral density $\sqrt{S_V}$ is measured as a function of the frequency for different $V_{pillar}$ values. An example is presented on Figure 1(b) corresponding to a device from stack 1 with $R_0$ = 45 kΩ. The red dashed line corresponds to the thermal noise $\sqrt{S_{V,th}} = \sqrt{4k_B TR}$ which varies from 3 to 30 nV/√Hz at room temperature depending to the TMR sensor resistances. The thermal noise is visible only for $V_{pillar} = 0$ mV, otherwise the shot noise should also be taken into account for the white noise background. Dashed black line is introduced to indicate the 1/f noise contribution. We do not observe random telegraphic noise (RTN) in our stack 1 devices[16] thanks to the free layer pinning[17]. So the 1/f low-frequency noise is the dominant contribution.

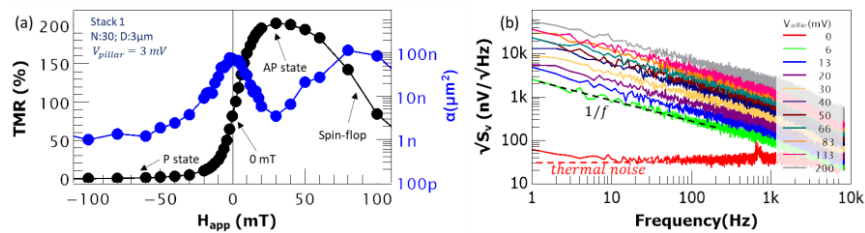

Fig. 1. (a) Magnetoresistance curve (black) and Hooge-like parameter α (blue) as a function of the applied field for one TMR magnetic sensor. (b) Noise amplitude spectral density as a function of the frequency and as function of the voltage bias. The cutoff frequency filter appears at 3 kHz. The two Figures correspond to the TMR of 30 pillars of 3 μm of stack 1.

First the 1/f low-frequency noise is quantified by extracting the parameter α as a function of the applied field, see Figure 1(a). The alpha parameter plotted is a mean value extracted between 1 Hz and 300 Hz. We first observe that $\alpha_P < \alpha_{AP} < \alpha_{0\,mT}$. On P and AP states, magnetic fluctuations are drastically reduced and the electric contribution of 1/f noise is dominant[18]. At 0 mT and also in the spin-flop region α takes a maximum values, one order of magnitude higher than in the P state, which is a characteristic of the 1/f low-frequency noise with a magnetic origin[19-23]. The electron tunneling in P and AP states are mainly dominated by different conductivity channels ($\Delta_1$, and $\Delta_2, \Delta_5, \Delta_{2'}$ respectively). Since $\Delta_1$ electrons present smaller parallel component of the wave vector ($k_\parallel$) in comparison with the non-localized $\Delta_2, \Delta_5, \Delta_{2'}$ electrons, they will be less affected by defects[24] (traps, vacancies, etc.) through the tunneling process given rise to $\alpha_P < \alpha_{AP}$[25, 26].

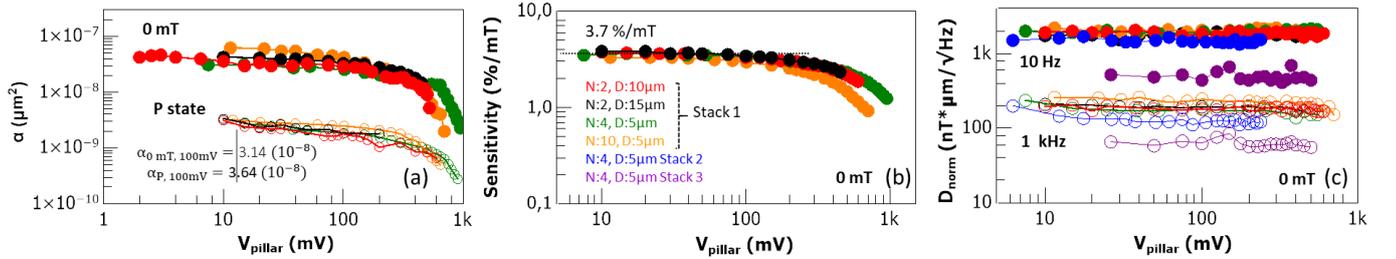

Fig. 2. Alpha parameter (a), sensitivity (b) versus voltage for 0 mT and P state for different devices of stack 1, 2 and 3. (c) normalized detectivity for several devices of the three different TMR stacks at 10 Hz and 1 kHz. The performances have been normalized and corrected as described in Ref 16.

Figure 2 shows the α parameter (a), the sensitivity (b) and the normalized detectivity (c) as a function of the bias voltage per MTJ pillar ($V_{pillar}$) for different TMR sensors at 0 mT. We observe how the 1/f low frequency noise quantified by α and the sensitivity decrease as a function of $V_{pillar}$. The measured detectivity shows a constant behavior versus $V_{pillar}$ highlighting an identical origin for sensitivity and noise decrease. The decrease of sensitivity might be understood by the reduction of magnetoresistance (MR) with voltage which is often explained by spin dependent scattering and symmetry filtering effect[14] in the case of TMR sensors. On the contrary, the abrupt α reduction at high voltages is more difficult to explain with the same mechanism. A first hypothesis is to relate the $\alpha$ variation to the non-linearity of the I-V responses. Indeed, Hooge's phenomenological formula describing the alpha parameter as constant with the applied voltage is only valid for elements with linear I-V responses, which is not the case for TMR elements. A small increase of the alpha parameter is observed at very low voltage bias and is probably an artefact due to the 1/f noise added by the amplifier which starts to be comparable to the 1/f noise due to the TMR.

The detectivity behavior versus voltage is observed for different devices and also for different TMR stacks and thus seems to be a general behavior for MgO TMR. The two TMR stacks with pinned free layer are stack 1 described previously and stack 2 which is similar except for the CoFe(1) which is replaced by a NiFe(10) layer and as well as a TMR stack 3 with a non-pinned free layer (composed only by CoFeB(1.5) / CoFe(1) / NiFe(6)). This constant detectivity, which has never been reported before, is a key feature for applications as it allows to control the bias voltage of this device without affecting the detection limit of the sensor.

**Model developed**

To explain the identical variation of the noise amplitude spectral density and the sensitivity and thus the constant detectivity, we introduce a model where each MTJ pillar of the TMR sensor is represented by three different transport contributions in parallel, as depicted in Figure 3(a). The first one is described by a 1/f noise spectral density $S_V$, a sensitivity $Sensitivity$ and a conductance $G_0$ at 0 mT, $G_P$ in the parallel state and $G_{AP}$ in the antiparallel state. Its sensitivity and conductance are considered independent of the bias voltage. This element is shortcut by an element having no noise, no sensitivity and showing a parabolic conductance behavior $G_S$. This second element represents the impact of the distortion of the MgO barrier shape as described by Brinkman and Simmons[27] and does not involve the different effects of the electronic transport from the different conductions channels Δ. The third contribution corresponds to the electronic transport from the different conductions channels $G_{CC} = f(V_{pillar})$, where f is an unknown function. This element has no noise and no sensitivity. In a first approximation, the effect of the conduction channels contribution is neglected and not considered for the model below.

The equivalent noise spectral density at 0 mT per pillar is computed using the Millman theorem[16] as follow:

$$\sqrt{S_{V,eq}}[V/\sqrt{Hz}] = \sqrt{\frac{G_0^2 S_V + G_S^2 S_S}{(G_0+G_S)^2}} = \frac{G_0\sqrt{S_V}}{G_0+G_S} = \sqrt{\frac{\alpha V_{pillar}^2}{A.f}} \cdot \frac{G_0}{G_0+G_S} = A'V_{pillar}\frac{G_0}{G_0+CV_{pillar}+DV_{pillar}^2} \quad (4)$$

The equivalent TMR is calculated as follows:
$$TMR_{eq} = 2.\frac{G_{P,eq}-G_{AP,eq}}{G_{P,eq}+G_{AP,eq}} = 2.\frac{(G_P+G_S)-(G_{AP}+G_S)}{\frac{G_P+G_{AP}}{2}+G_S} = 2.\frac{G_P-G_{AP}}{G_0+G_S} = 2.\frac{G_P-G_{AP}}{G_0}\cdot\frac{G_0}{G_0+G_S} = 2.TMR.\frac{G_0}{G_0+G_S} \quad (5)$$

So the sensitivity can be estimated as the $TMR_{eq}$ ratio divided by the saturation field and multiplied by the applied voltage:

$$Sensitivity_{eq}\left[\frac{V}{T}\right] = \frac{TMR_{eq}.V_{pillar}}{2B_{Sat}} = \frac{V_{pillar}}{2B_{Sat}}.2.TMR.\frac{G_0}{G_0+G_S} = 2.Sensitivity\frac{G_0}{G_0+G_S} = A''V_{pillar}\frac{G_0}{G_0+CV_{pillar}+DV_{pillar}^2} \quad (6)$$

The equivalent conductance is the sum of the conductance of each contribution:
$$G_{eq} = G_0 + CV_{pillar} + DV_{pillar}^2 \quad (7)$$

The equivalent noise, sensitivity and conductance are estimated for one pillar in function of $V_{pillar}$. The contribution from the N pillar has to be taken into account for comparison with the experimental measurements (equations 1, 2 and 3). The total noise is equal to $\sqrt{N}$ times equivalent noise per pillar, and the total sensitivity is the mean of the sensitivity equivalent per pillar[16] Thus for N identical pillars: $\sqrt{S_{V,total}}[V/\sqrt{Hz}] = \sqrt{N.S_{V,eq}}[V/\sqrt{Hz}]$, $Sensitivity_{total}[\%/mT] = Sensitivity_{eq}[\%/mT]$ and $G_{total} = G_{eq}/N$.

The fits on the measured noise amplitude spectral density and sensitivity are presented in Figure 3(b) and (c) using expressions (4) and (6) and show the same parameters $G_0 = 51.4$, $C = -50$ and $D = 300$ in both cases. They are in good agreement with the experimental data up to $0.6\ V$ per pillar. From equations (3), (4) and (6), the detectivity value is calculated, finding that it corresponds to the ratio $\frac{A'}{A''} = 138\ nT/\sqrt{Hz}$. It is possible to normalized the detectivity multiplying by the factor $\sqrt{N.A}$[16], giving rise to a detectivity of $1.93\mu T.\mu m/\sqrt{Hz}$ in good agreement with the experimental of $2.04 \pm 0.02\ \mu T.\mu m/\sqrt{Hz}$, see Figure 2(c). Surprisingly the sensitivity, noise and detectivity dependence versus voltage could be explained without taking into account the effect of conduction channels represented by $G_{CC}$, which is not the case for the TMR conductance versus $V_{pillar}$, as it is shown in Figure 4.

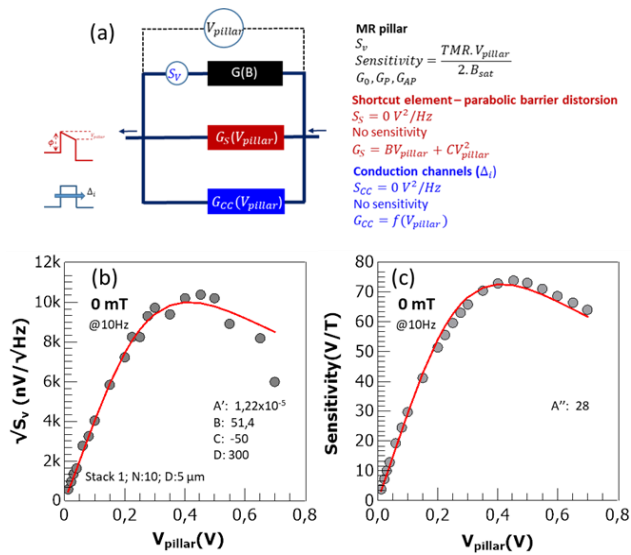

Fig. 3. (a) Schematic of the model proposed to explain the sensitivity and noise amplitude spectral density variation versus bias voltage: each TMR pillar is shortcut by two elements having a conductance $G_S(V_{pillar})$ and $f(V_{pillar})$ without noise contribution. (b) Noise amplitude spectral density at 10 Hz, (c) sensitivity as a function of the bias voltage by pillar for 0 mT for the devices (10 pillars of 5 μm) of stack 1. Experimental data are presented by dots, the model fit with a red line.

## Non linear I-V curves: signature of conduction bands

As it is well known, in CoFe/MgO/CoFe MTJ structure the conductance is the sum of several contributions of majority and minority electrons with the corresponding delta symmetries ($\Delta_1, \Delta_2, \Delta_5$ and $\Delta_{2'}$). These different symmetries of conduction channels have been detailed in[11, 14] and presented schematically in Figure 4(a) are related to the conductance anomalies at $0.3 - 0.35\ V$ and above 0.6 V. These anomalies are reflected in changes in the trend of conductance (Figure 4(b)) and correspond to the opening or closing of conduction channels with different symmetry and depending on the applied voltage per pillar.

Following the previous model, the total conductance should be $G_{total} = \frac{G_{eq}}{N} = E \cdot \frac{G_0 + CV_{pillar} + DV_{pillar}^2}{N}$ where $E$ is a constant factor ($E = 2.10^{-6}$). The curve is shown as a grey dash line in Figure 4(b). As expected this model does not explain the experimental conductance which clearly shows discrepancy from the parabolic fit and highlights the impact of the conductance channels, $G_{CC}$, which has to be taken into account. A parabolic fit is done in the voltage range between $0.5 - 0.7\ V$, follows the Brinkman model and allows to extract the barrier height $\phi$ and the asymmetry $\Delta\phi$: $\phi = 1.83\ eV, \Delta\phi = -0.052\ eV$ and $G_{0\ mT} = 0.173\ mS$. These values are consistent with typical MgO tunnel junctions. This fit is shown as a red line in Figure 4(b). The difference between the measured conductance and the parabolic fit highlights the conductance anomalies (blue squares) associated with the opening or closing of conduction bands as described in the schematic of Figure 4. More precisely a combination of the P and AP state should be taken into account as the conductance is measured at 0 mT. Thus, even if the TMR conductance has a typical behavior and emphasizes the contribution of different bands, the noise and sensitivity variation versus voltage can be explained without.

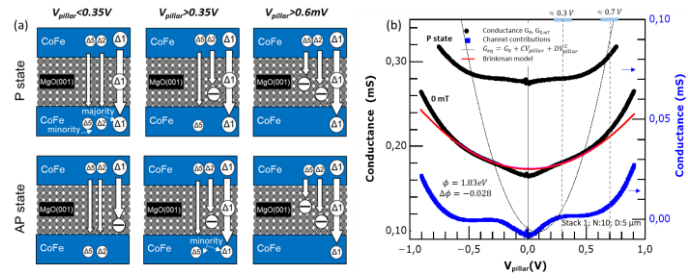

Fig. 4. (a) Schematic of the majority and minority symmetry electrons tunneling present for each voltage range in the P and AP state. (b) Conductance versus voltage in the P state and at 0 mT for the device (10 pillars of 5 µm) of stack 1. The fits represent the parabolic behavior of a corrected shortcut element (red curve). The blue squares dots are the difference between the conductance measured and the parabolic fit and highlight the channel contributions.

## Frequency and high voltage behavior

We then investigate the limits of the model, measuring the noise and the detectivity at high bias voltages and plotting the measurements versus frequency, see Figure 5(a). A discrepancy of the model for voltage higher than $0.6\ V$ appears and seems to be more pronounced at low frequency. Several origins of this discrepancy are possible. The elevation of temperature induced by joule heating at high voltage could be responsible of sensitivity change. A spin diffusion increase present at high voltage[28] and also seen in the TMR ($V_{TMR}$) curve might also be responsible for the change of sensitivity decrease. The band transitions at $0.6\ V$, localized states or traps inside the barrier might also play a role[29-31]. The presence of a RTN at low frequency could potentially explain the deviation to the model at low frequency. Moreover at high voltage, the conductance might not be parabolic but this could not explain the discrepancy of the model. The diffusion of atoms inside the barrier due to electric field might be ruled out because the effect observed is reversible.

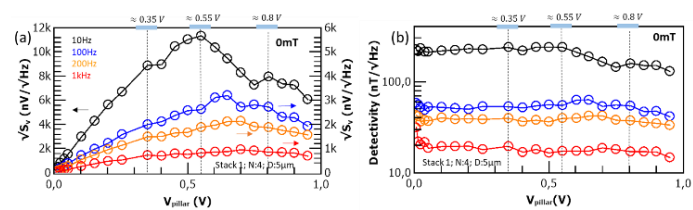

Figure 5. Noise amplitude spectral density (a) and detectivity (b) of a device (4 pillars of 5µm) of stack 1 as function of voltage bias up to 1 V and of the frequency (10 Hz to 1 kHz).

## Conclusions

In this work, we demonstrate experimentally that $\alpha$ factor and the sensitivity present an unexpectedly decrease with the bias voltage inducing a surprising constant detection limit. The constant detectivity is experimentally observed for several devices and TMR sensor stacks and thus presents a universal character. With a model in which the TMR sensor is shortcut by an element having no noise and a parabolic variation versus voltage, we can describe the distortion of the MgO barrier versus voltage and reproduce the experimental behaviour up to a voltage per pillar of 600 mV whereas the TMR sensor conductance highlight the transport through typical conductions bands.

## Acknowledgements

This work was supported in part by the French ANR under Projects AdvTMR (ANR-18-ASTR-0023), NeuroTMR (ANR-17-CE19-0021) and CARAMEL (ANR-18-CE42-0001).

## References

1. Kent, A., Worledge, D. A new spin on magnetic memories. Nature Nanotech 10, 187–191 (2015).
2. Zeng, Z., Finocchio, G., Zhang, B. et al. Ultralow-current-density and bias-field-free spin-transfer nano-oscillator. Sci Rep 3, 1426 (2013).
3. Torrejon, J., Riou, M., Araujo, F. et al. Neuromorphic computing with nanoscale spintronic oscillators. Nature 547, 428–431 (2017).
4. C. Huang, C. Lin, J. Kao, J. Chang, and G. Sheu, IEEE Trans. Veh. Technol. 67, 6882 (2018).
5. R. Guerrero, M. Pannetier-Lecoeur,C. Fermon, S. Cardoso, R. Ferreira and P. P. Freitas, J. Appl. Phys. 105,113922 (2009)
6. R. C. Chaves, P. P. Freitas, B. Ocker, and W. Maass J. Appl. Phys. 103, 07E931 (2008)
7. D. Leitao et al. IEEE Transactions on Magnetics, Vol. 48, No. 11, (2012)
8. E.R. Nowak, P. Spradling, M.B. Weissman, S.S.P. Parkin. Thin Solid Films 377-378 , (2000), 699-704
9. Yuasa, S., Nagahama, T., Fukushima, A. et al. Giant room-temperature magnetoresistance in single-crystal Fe/MgO/Fe magnetic tunnel junctions. Nature Mater 3, 868–871 (2004)
10. S. S. P. Parkin et al., Nature Mater. 3, 862 (2004)
11. S. Ringer, M. Vieth, L. Bar, M. Ruhrig and G. Bayreuther, Phys. Rev. B 90, 174401 (2014)
12. E. Monteblanco, F. Garcia-Sanchez, D. Gusakova, L. D. Buda-Prejbeanu, U. Ebels, J. Appl. Phys, 121, 013903 (2017)
13. D. C. Leitao, A. V. Silva, R. Ferreira, E. Paz, F. L. Deepack, S. Cardoso, and P. P. Freitas, J. Appl. Phys. 115, 17E526 (2014)
14. William H Butler, Sci. Technol. Adv. Mater. 9 (2008) 014106 ; W. H. Butler et al. PRB, 63 054416 (2001)
15. Ana V. Silva, Diana C. Leitao, Joao Valadeiro, Jose Amaral, Paulo P. Freitas, and Susana Cardoso, Eur. Phys. J. Appl. Phy. 72: 10601, (2015)
16. Paper IEEE Sensors E. Monteblanco, Submitted (2020).
17. J. Moulin, A. Doll, E. Paul, M. Pannetier-Lecoeur , C. Fermon, N. Sergeeva-Chollet, and A. Solignac, Appl. Phys. Lett. 115, 122406 (2019)
18. A. Faridah Md Nor, T.Kato, S. Jin Ahn, T. Daibou, K. Ono, M. Oogane, Y. Ando, and T. Miyazaki J. Appl. Phys 99, 08T306 (2006)
19. Z. Diao, J. F. Feng, H. Kurt, G. Feng, and J. M. D. Coey, APL 96, 202506 (2010)
20. J. M. Almeida, P. Wisniowski, and P. P. Freitas IEEE Trans Mag, Vol. 44, No. 11, (2008)
21. Aisha Gokce, E. R. Nowak, See Hun Yang, S. S. P. Parkin, J. Appl. Phys. 99, 08A906 (2006)
22. F. G. Aliev, et al. Appl. Phys. Lett. 91, 232504 (2007)
23. C. Fermon, M. Pannetier-Lecoeur, Noise in GMR and TMR Sensors, Giant Magnetoresistance (GMR) Sensors: From basics to State of the art Applications. Springer, Berlin, Heilderberg, 2013, pp. 47-70
24. Z. Diao, J. F. Feng, H. Kurt, G. Feng, and J. M. D. Coey, APL 96, 202506 (2010)
25. J. Scola, H. Polovy, C. Fermon, M. P. Lecoeur, G. Feng, K. Fahy, and J. M. D. Coey. Appl., Phys. Lett., vol. 90, p. 252501, (2007)
26. Ryan Stearrett, W. G. Wang, L. R. Shah, Aisha Gokce, J. Q. Xiao, and E. R. Nowak, J. Appl. Phys. 107, 064502 (2010)
27. W. F. Brinkman, R. C. Dynes, and J. M. Rowell, J. Appl. Phys 41, 1915 (1970)
28. Spin diffusion reference
29. B. Taudul et al. Adv. Electron. Mater. 2017, 3, 1600390
30. J. M. Teixeira, J. Ventura, M. P. Fernández-García, J. P. Araujo, J. B. Sousa, P. Wisniowski, and P. P. Freitas, Appl. Phys. Lett. 100, 072406 (2012)
31. F. G. Aliev, J. P. Cascales, A. Hallal, M. Chshiev, S. Andrieu, PRL 112, 216801 (2014)